# $Nd_{2-x}Ce_xCuO_{4-y}/Nd_{2-x}Ce_xO_y$ boundary and resistive switchings in mesoscopic structures on base of epitaxial $Nd_{1.86}Ce_{0.14}CuO_{4-y}$ films

N.A.Tulina[1], A.N.Rossolenko[1], A.A.Ivanov[3], V.V.Sirotkin[2], I.M.Shmytko[1], I.Yu.Borisenko[2], A.M.Ionov[1]

[1]Institute of Solid State Physics RAS, Chernogolovka, tulina@issp.ac.ru
[2]Institute of Microelectronics Technology and High Purity Materials RAS, Chernogolovka
[3]National Research Nuclear University "MEPhI", Moscow, Russia

*Reverse and stable bipolar resistive switching effect (BRSE) was observed in planar $Nd_{2-x}Ce_xCuO_{4-y}/Nd_{2-x}Ce_xO_x/Ag$ heterostructure. It was shown that the CVC of the BRSE observed has a diode character. Simulations were used to consider the influence of the nonuniform distribution of an electric field at the interface of a heterojunction on the effect of bipolar resistive switching in investigated structures. The inhomogeneous distribution of the electric field near the contact edge creates regions of higher electric field strength which, in turn, stimulates motion and redistribution of defects, changes of the resistive properties of the whole structure and formation of a percolation channel.*

## 1. Introduction.

Resistive switching (RS) memory is now one of the most promising types of memory based on novel physical principles [1-7]. The resistive switching effect is observed in a wide range of structures with a dielectric layer composed of simple oxides as well as complex compounds. The surge in research in 2008 was stimulated by the HP studies proposing the use of the RS effect for creating a memristor, a fourth basic element along with resistor, capacitor and inductor, i.e. for creating logical chains and circuit design [8]. The research and publication activity didn't grow down so far. Review [9] classifies the studies of RS in nearly 100 different double oxides, complex oxide compounds and other types of materials. The microscopic pattern of diverse factors that are essential and sufficient for reproducible and stable switching is yet to be established. Neither model is capable of complete explaining the RS phenomenon owing to lack of fundamental understanding of the RS mechanism at the atomic scale, i.e. the mechanism of formation and disintegration of conductive paths.

All RS transitions under study can be divided into structures with a vertical and planar geometry. The first type is represented by symmetric micro- or nano- sized metal-insulator-metal (MIM) transitions. Planar structures are also MIM transitions, but their micro-size is formed by the upper point-like electrode. In this case the RS is significantly affected by the distribution topology of the electric field [10, 11]. It was also demonstrated that the essential condition for observation of the RS effect is the presence of a surface layer (carrier-depleted layer) of sizes of the order of 10 nm with conductivity different from bulk conductivity. The bipolar resistive switching effect (BRSE) observed in heterostructures based on oxide compounds including those based on strongly correlated electronic systems (SCES) such as high temperatures superconductors HTSC (see, for instance, review [7]) appeared to be an original method sensitive to SCES-type carriers. At certain polarity of the electric field the phase composition of the SCES surface layer changes at the nano-scale level. This realizes metastable high resistance and low resistance states (Off and On, respectively) of the heterocontact and induces colossal electroresistance (CER) [7] which is an $R_{Off}/R_{On}$ resistance ratio characterizing the memory effect. Studies of BRSE in different oxide compounds revealed that a role of oxygen diffusion is significant, although not exclusive. BRSE realizes metastable states with different oxygen contents in the interface region in the compound in question. The resistive properties of the structure are determined by that particular region and, hence, BRSE induces reproducible oxygen doping of a certain sample region.

$Nd_{2-x}Ce_xCuO_{4-y}$ occupies a special place among copper-based oxide superconductors with a perovskite structure. Standard HTSC materials contain conducting $CuO_2$ layers with oxygen pyramids (YBCO, BSCCO) or octahedrons (LSCO), while optimally annealed $Nd_{2-x}Ce_xCuO_4$ crystals contain $CuO_2$ layers without apical oxygen atoms. $CuO_2$ layers form quasi-two-dimensional (2D) planes spaced at $a = 6$ A. Therefore, $Nd_{2-x}Ce_xCuO_4$ single crystals can be considered as a selectively doped system of quantum wells ($CuO_2$ layers) separated by cerium-doped barriers (buffer NCO layer). This statement is based on the fact that layered HTSC materials demonstrate well-pronounced 2D carrier properties in macroscopic 3D crystals. Undoped $Nd_2CuO_4$ is a dielectric. Cerium-doping of $Nd_2CuO_4$ and lowering of the oxygen content to the stoichiometric level results in n-type conductivity in the $CuO_2$ layers of the $Nd_{2-x}Ce_xCuo_4$ crystal. As extra electrons escape to the $CuO_2$ plane, it creates a potential of charged $Ce^{4+}$ impurity centers randomly distributed in the lattice. The scattering on the potential is likely to determine carrier mobility.

Previously [12] we obtained $Nd_{1.86}Ce_{0.14}CuO_{4-y}$ (NCCO) films with an ingrown second dielectric $Nd_{0.5}Ce_{0.5}O_{1.75}$ (NCO) phase. Several works (performed on single crystals and HTSC films) revealed such epitaxial growth of the second phase [13]. Thus, such film structures are *unique objects* for studies of BRSE when the buffer layer is an oxygen- and cerium-doped dielectric and the base NCCO film is a metal (superconductor) and the NCCO/NCO boundary is the internal barrier for $Nd_{2-x}Ce_xCuO_4$ crystals. This work presents the results of the investigation of BRSE in NCCO/NCO-based structures

## 2. Experimental

NCCO films were obtained by the pulse laser deposition technique. The substrates were $SrTiO_3$ (100) single crystals 5x10 mm$^2$ in size. The quality of the films was examined by x-ray photoelectron spectroscopy and x-ray analysis using a D500 (Siemens) diffractometer. The topological structure of the film surface was studied using a SUPRAII scanning microscope and a Solver PROM (NTMDT) atomic-force microscope. The phase composition and crystalline structure of the film were studied by Xray diffraction in the Bragg–Brentano geometry using monochromatic Cu$K\alpha$ radiation. A set of (00l) reflections belonging to the NCCO phase; and a set of (h00) reflections corresponding to the NCO phase were detected (Fig1). Both phases were found to be epitaxial.

Two types of mesoscopic NCCO/NCO/Ag heterostructures were used: the first was of a microcontact type, the other was created by means of photolithography. The microcontact junctions were produced by bringing the metal needle of the micromanipulator to the as-grown film surface. The lithography-formed NCCO/NCO/Ag structure had a contact window 50x50 mkm$^2$ in size. The current lead was made by pulsed laser deposition of Ag film on a side structure surface, as shown in Figs.1,3.

The heterojunctions were studied in order to reveal resistive switching by measuring the current-voltage characteristics (CVC) and the temperature dependence of the heterocontact metastable states resistance. Diverse metastable states were realized by varying the external parameters, namely, the frequency and voltage of the electric field applied to the heterocontact.

The following conditions should be met to realize resistive switching: the resistivity of the base contact should be considerably lower than that of the interface, and the latter should contain a system of mobile charge carriers. It should also be emphasized that it is the resistive properties of the interface that must determine the total resistance of the heterostructure. Resistivity of copper-based HTSCs is highly anisotropic [14, 15]. The films oriented along the <ав> plane exhibit a quadratic resistivity-temperature dependence characteristic of optimally doped $Nd_{1.85}Ce_{0.15}CuO_{4-y}$. Those oriented along the <ac> and <bc> planes demonstrate a more complicated behavior and significantly higher resistivity. For this reason only (001) oriented films are suitable for BRSE observation. Note that RS was observed after preliminary training of the structure under negative voltage applied to bottom electrode (NCCO film Fig.1). This phenomenon is known as "electroforming".

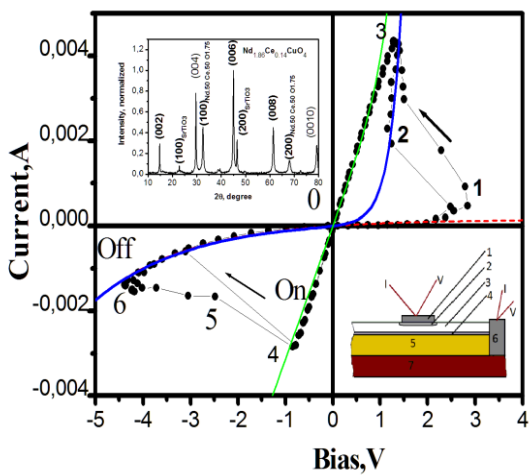

*Fig.1. Examples of CVC of RS observed in Ag/NCO/NCCO heterostructures and of the electroforming process in Ag/NCO/NCCO heterostructures. The solide lines indicate the CVC approximations by ratios 1 (see the comments in the text). The x-ray spectrum is shown in the left upper corner. The structure is schematically shown in the lower right corner: 1. Metallic electrode, the diameter d. 2. Depleted region on the metal-NCO semiconductor interface (Schottky barrier). 3. Current spread region in NCO film, h=30nm. 4. Depleted region on the NCO (semiconductor )/ (NCCO) (superconductor ) interface (p-n junction or the Schottky barrier). 5. Superconducting NCCO film, $T_c$ = 21.5 K, H=300nm. 6. The Ag film was made by the pulse laser deposition technique on a side structure surface for current lead. 7. STO substrate.*

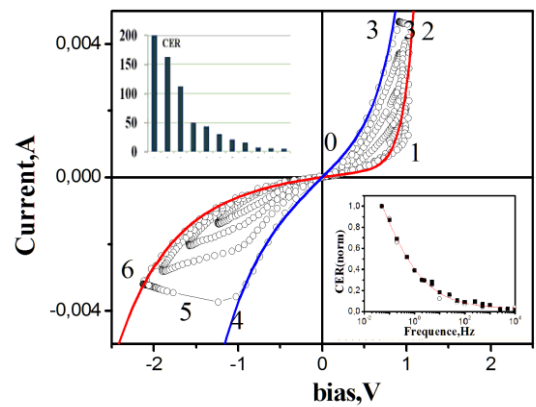

*Fig.2. Example of CVC of metastable states-exhibiting NCCO/NCO-based heterostructure with an upper Ag microcontact electrode. The dashed lines indicate the CVC approximations by ratios 1. (see the comments in the text). Upper left corner: CER of various metastable states NCCO/NCO/Ag heterocontacts. Lower right corner- CER normalized on the maximum slow scan value as a function of applied voltage frequency in the NCCO/NCO/Ag heterocontacts.*

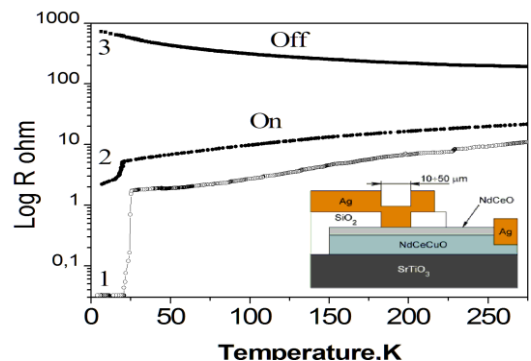

*Fig.3. Temperature dependence of resistance of NCCO film (1) and metastable ON(2) and OFF (3)*

*states. Lower right corner: schematic of lithography-formed structure representation*

Fig.1, 2 show examples of the CVC of the RS observed in the heterostructures. The high-resistivity branch of the CVCof the heterocontacts has a diode character. Switching starts at a certain positive threshold voltage at top electrode (Ag) .The CER value is a function of the contact area, the thickness of the dielectric layer NCO, value of the applied voltage and that of the limited current (Fig.2) . The best reproducible results were obtained to 100 cycles on the microcontacts. We have also investigated changes of the electrodynamic properties of the heterojunctions exposed to a sinusoidal ac electric field by measuring the CVC of the heterojunctions at $10^{-3}$ – $10^5$ Hz and recording the current-voltage curves of the heterocontact. Fig.2 shows CER normalized on the maximum slow scan value as a function of applied voltage frequency in the NCCO/NCO/Ag heterocontacts. The increase in the applied voltage frequency involves an increase in the switching voltage and a decrease in the resistivity changes in the on- and off-states (CER), which results in collapse of the on- and off-state branches. Therefore, frequencies of the order of $10^4$ – $10^5$Hz are threshold for observation of BRSE in the structures under study.

To explain the switching mechanism let us consider the heterojunction components and how they are affected by the electric field. The high-resistivity branch (6-0-1 in Fig.1-2) of the CVC of the BRSE-exhibiting heterocontacts shows a diode character that enables us to assume that they are Schottky-like diodes. The branch (6-0-1-2) can be approximated by current transport of two oppositely parallel connected diodes (Fig.2) . The CVC of the structures suggests the presence of parallel short-circuits $R_i$, which should be taken into account when modeling the transport properties of the structures by the relationship:

$I = -I_{01} *(\exp((-V)/n_1*kT)-1) + I_{02}*(\exp((V)/n_2*kT)-1)+V/R_i$, (1)

where $I_0 = A^*T^2*S*\exp(-(e\phi_B)/kT)$, $A^*$ is the Richardson constant , n is the imperfection factor, $\phi_B$ is the barrier height, S is the contact area. In heavily doped defect structures the height and width of the barrier depend on its charge system. Transition from the high resistance to the low resistance state (and backwards) is determined by formation and disintegration of the percolation path on CVC branches "1-3", "4-6".

**3.Computational simulation (Model of the critical region)**

The physical processes induced by BRSE can be described using numerical simulation. Two factors underlie the RS process: the effect is determined by the electric field forming a percolation path and a reverse process, i.e. disintegration of the conducting channel. Developing the proposed method of numerical calculation of the resistive properties of heterocontacts based on oxide compounds [16,17], we calculated resistances $R_{On}$, $R_{Off}$, the equipotential and current distributions upon electric field sweep in the interface of the heterocontacts in question. In the model of the "critical region", when the formation mechanism of the conducting channel is not specified, it is assumed that in the interface region the electric field is inhomogeneously distributed, which involves formation of zones with a strong local increase of the electric field strength. This may be due to the influence of the electrode shape (considered here) or structural, or technological inhomogeneity. In this work we study the case of a point heterocontact (Fig.4). This model is a development of the previously proposed one [17] and takes into account anisotropy of the resistive properties of the heterostructure.

The mathematical description of the model is based on the Poisson's equation

$$\nabla[\sigma(r, z)\nabla\varphi(r, z)] = 0; \qquad (2)$$

that enables to determine potential distribution $\varphi$ and then field distribution (strength) by the resistivity distribution in the heterostructure:

(1) The distribution of the electric potential $\varphi$ is calculated from the distribution of specific conductivity $\sigma$ in the structure.

(2) The distribution of electric field intensity $E$ is found;

(3) If there are areas in the defective layer in which $E \geq E_{cr}$, $\sigma_{on}$ is replaced by $\sigma_{off}$ and the procedure returns to step (1).

The numerical realization of the model involved a combination of an integro-differential approximation and a multi-grid approach [16], which provided an efficient calculation tool for analysis of the processes of formation and disintegration of a "conducting channel". Fig.4 demonstrates results of simulation as step-by-step change of isopotential map and current lines distribution for different CVC segments of the RS-exhibiting heterocontacts, the arrows indicate the sweep direction. The high resistance state is determined by the properties of the structure, the CVC branch {6-0-1}. Carrier tunneling starts at point 1, the properties of the heterostructure are changed, as a result there are two channels for current transport, which can be approximated by current transport of two oppositely parallel connected diodes and (in this case) the resistance is reduced (branch 1-3) by an order of magnitude and a ring shaped conductive channel form with transverse sizes of a radius r, $\Delta r$, and h~30 nm. The low resistance state -branch 3-4, branch 4-6 exhibits transition to the high resistance state. Branches 6-1 and 3-4 are reversible, switching occurs on branches 1-3 and 4-6.

Reverse voltage polarity induces the process of disintegration of the "conducting channel". The allowance made for the anisotropy of the superconducting NCCO film yields a result: a high local strength region occurs on the NCCO/NCO interface, and it is in that region where disintegration of the conducting channel proceeds (Fig.4c).

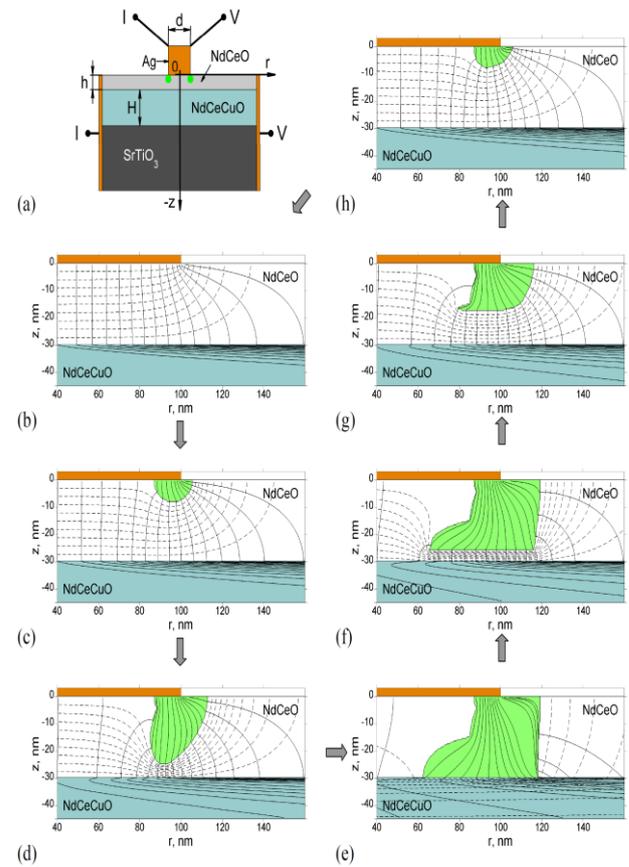

*Fig.4. Correlation between the CVC branches of the heterocontacts (Fig1) and the calculated model. In the pictures is also shown distribution of equipotential (dashed) and current (solid) line. (a) Schematic drawing of the structure. (b-h) The evolution of conducting channel : (b) OFF-state {6-0-1}, (c-d) formation of conductive channel {1-2-3}, (e) ON-state (the conductive cannel in the form of the ring with cross sizes r, Δr, and h ~30 nm.){3-4}, (f-h) decomposition of conductive channel {4-6}. The numbers in curly brackets refer to the corresponding branch of the current-voltage characteristic curve from Fig. 1.*

The formation of a percolation channel during the initial electroforming process creates a certain distribution of high and low conductivity regions, their ratio being responsible for the formation of a total resistance and creating conditions for reproducible switching. As shown by the calculations, such regions are created by the edge effect of planar heterosructure.

### 4. Discussion

The mechanism of the influence of electric field on the NCO resistive state is still an unresolved issue. $Nd_{0.5}Ce_{0.5}O_{1.75}$ is an oxygen deficient oxide. Oxides of transition and rare earth (Re) metals reveal a significant deviation from stoichiometry caused either by an oxygen or metal deficiency. The effect of oxygen deficiency on the resistive properties of oxides is commonly known. The conductivity of such oxides is ionic and proceeds through a vacancy mechanism [18]. The ability of Re metals to change their valence state, easy oxygen exchange with the environment, diversity of defects forming discrete levels in the forbidden gap have a considerable effect on the properties of the compounds in question. Let us consider two feasible variants.

The first is the Poole - Frenkel effect when the electric field $E>E_{rh}$ increases the number of carriers on the metal-NCO interface (branche 1-2), which induces current tunneling and formation of a percolation path. As a result there are two channels of conductivity in the contact area, which can be approximated as two parallel diodes. The manifestation of the Poole-Frenkel mechanism in the course of current transfer points to the presence of localized donor centers. Formation of a percolation channel proceeds as follows: the increasing electric field induces emission of a free charge from the defect accompanied by redistribution of the electric field, its local increase between the regions with a high free charge density, which stimulates local emission, field redistribution and, finally, leads to formation of a percolation path. The main mechanisms affecting channel formation are free charge emission from the existing centers, on the one hand, and formation of new metastable centers in the electric field, on the other hand. It is shown that defect oxides contain a number of such traps related to bond cutting or rearrangement (the so-called alternative valence defects resulting from simultaneous occurrence of equal concentrations of donor and acceptor defects). Donor centers (oxygen vacancies) generate Poole - Frenkel conductivity while acceptor traps (probably $Ce^{4+}$) are paths for hopping conductivity. Work [19] considers the RS mechanism generated by electron transition conditioned, in turn, by ionization of traps by the Poole-Frenkel mechanism and metallization of the semiconductor in which the critical carrier concentration, the so-called concentration instability, is reached.

Another variant is that such a strong electric field stimulates also electrodiffusion of mobile defects (oxygen vacancies) and, hence, allows modulating the barrier. In the case in question there is also an internal barrier, which is essential for the reverse and polar switching effect. In this respect, the observed frequency dependence of BPRS confirmed the significant role played by the diffusion processes. As it is seen from Fig.2, the frequency of applied voltage of the order of $10^4$ Hz is a threshold value for BPRS observation. Thus, the frequencies suppressing the switching effect in our structures are fairly low, which is indicative of a diffusion mechanism. The frequency characteristics suggest a memristor nature of the heterojunctions [12]. It is quite obvious that RS polarity and reproducibility cannot be explained without taking account of ellectrodiffusion that changes the charge state in the active transition region. To achieve RS reproducibility, the structure should satisfy a topological requirement: the distribution of the electric field should ensure an electric field in the key voltage sweep points which is sufficiently strong for oxygen electrodiffusion. The field behavior of BRSE is indirectly confirmed by the observation of the higher switching voltage in the structures with thicker NCO films.

The temperature dependence of metastable state resistivity $R_{on}$ allows calculating the contributions of the structure components to the resistance. Fig.3 shows that, first, the temperature dependence of metastable on state resistance corresponds to the behavior of thin NCCO films Fig.3 (curve 1) and exhibits two characteristic components: the residual resistance of the percolation channel $R_{o1}$ and $R_{o2}$ - the resistances resulting from the transition of the NCCO film contacts to the superconducting state at $T=T_c$. Given all the parameters of the heterocontacts in our case, comparison can be made between the estimations of the changes of the resistive properties and the calculated values (Table 1).

Table 1. Examples of experimental and calculated data for the heterojunctions in question.
$R_{off}$, $R_{on}$ - the resistancies of heterostructure metastable states; $r_0$ - the radius of top Ag electrode; $\Delta r$ - the thickness of the conductive cannel, $\rho_{off}$, $\rho_{on}$ – the resistivity of the conductive channel.

|    | $R_{off}$ ohm | $R_{on}$ ohm | CER | $r_0$ cm | $\Delta r_0$ cm | $R_{off}$ cal ohm | $R_{on}$ cal ohm | $\rho_{off}$ cal Ohm*cm | $\rho_{on}$ cal Ohm*cm |
|----|--------|-------|-----|---------|---------------|-----------|----------|----------------|---------------|
| #1 | 43380  | 381   | 114 | $1*10^{-5}$ | $5.55*10^{-6}$ | 43380 | 380 | 5.8 | 0.01 |
| #2 | 33587  | 326   | 103 | $1.2*10^{-5}$ | $4.76*10^{-6}$ | 33589 | 324 | 6.2 | 0.01 |
| #3 | 12764  | 195   | 65  | $2.1*10^{-5}$ | $0.82*10^{-6}$ | 12760 | 196 | 6.65 | 0.0098 |
| #4 | 10000  | 640   | 15  | $1*10^{-5}$ | $4,61*10^{-6}$ | 10002 | 640 | 1.3 | 0.024 |
| #5 | 2314   | 387   | 6   | $1*10^{-5}$ | $3.24*10^{-6}$ | 2314 | 386 | 0.28 | 0.004 |
| #6 | 440    | 24    | 18  | $2,5*10^{-3}$ | $2.5*10^{-3}$ | 440 | 24 | 2723 | 0.05 |
| #7 | 448    | 33    | 14  | $2.5*10^{-3}$ | $0.08*10^{-4}$ | 448 | 33 | 2779 | 0.17 |
| #8 | 100    | 27    | 4   | $2.5*10^{-3}$ | $2.5*10^{-3}$ | 101 | 27 | 580 | 0.24 |

It is seen that the agreement between the observed and calculated values is fairly reasonable. The calculated values of resistivity for the conducting channel are close to limit which can are observed in heavily doped semiconductors and don't exclude metal inclusions in the form of for example trimers {Nd/CeVo¨Nd/Ce}, where Vo¨ - oxygen vacancy [18].

**5. Conclusion**

Thus, the crucial factor for BRSE observation is the presence of an in-grown second epitaxial oxygen–deficient NCO phase on the NCCO surface. The inhomogeneous distribution of the electric field near the electrodes (Ag and NCCO) creates regions of higher electric field strength, which, in turn, stimulates motion and redistribution of defects in the NCO film and changes the resistive properties of the whole structure. The resistance switching in Schottky junctions arises through emergence additional tunneling paths rather than the change in barrier potential profile. The basic factors are the presence of oxygen vacancies in the NCO phase, the geometric size of the structure and existence of barriers on the structure electrodes (top-Ag/NCO, bottom-interface NCO/NCCO).

**Acknowledgments**

This work was supported by the Russian Foundation for Basic Research (grant No, 14-07-00951) and the programs of the Physical Sciences Division RAS "Physics of Novel Materials and Structures".